\documentclass[11pt,preprint]{aastex}

\def\tcorner{$t_{\rm corner}$}
\def\bprojected{$B_{\rm proj}$}
\def\bunprojected{$B_{\rm total}$}
\def\srms{S_{\rm rms}^2} 
\def\nrms{N_{\rm rms}^2} 
\def\nrmsf{N_{\rm rms}^4} 

\title{A Multi-Baseline 12 GHz Atmospheric Phase Interferometer with
  One Micron Path Length Sensitivity}

\author{Robert S. Kimberk\altaffilmark{1}, 
  Todd R. Hunter\altaffilmark{2}, 
  Patrick S. Leiker\altaffilmark{1}, 
  Raymond Blundell\altaffilmark{1},
  George U. Nystrom\altaffilmark{1},
  Glen R. Petitpas\altaffilmark{3},
  John Test\altaffilmark{1}, 
  Robert W. Wilson\altaffilmark{1}, 
  Paul Yamaguchi\altaffilmark{3},
  Kenneth H. Young\altaffilmark{1}, 
  }

\email{thunter@nrao.edu}
\altaffiltext{1}{Harvard-Smithsonian Center for Astrophysics, 60 Garden Street, Cambridge, MA 02138} 
\altaffiltext{2}{National Radio Astronomy Observatory, 520 Edgemont Road, Charlottesville, VA, 22903} 
\altaffiltext{3}{Submillimeter Array, 645 North A'ohoku Place, Hilo, HI 96721} 
 
\begin{document}

\begin{abstract}

We have constructed a five station 12~GHz atmospheric phase
interferometer (API) for the Submillimeter Array (SMA) located near
the summit of Mauna Kea, Hawaii.  Operating at the base of unoccupied
SMA antenna pads, each station employs a commercial low noise mixing
block coupled to a 0.7 m off-axis satellite dish which receives a
broadband, white noise-like signal from a geostationary satellite.
The signals are processed by an analog correlator to produce the phase
delays between all pairs of stations with projected baselines ranging
from 33--261 m.  Each baseline's amplitude and phase is measured
continuously at a rate of 8 kHz, processed, averaged and output at 10
Hz. Further signal processing and data reduction is accomplished with
a Linux computer, including the removal of the diurnal motion of the
target satellite.  The placement of the stations below ground level
with an environmental shield combined with the use of low temperature
coefficient, buried fiber optic cables provides excellent system
stability.  The sensitivity in terms of rms path length is 1.3
microns which corresponds to phase deviations of about 1\arcdeg\
of phase at the highest operating frequency of the SMA.  The two
primary data products are: (1) standard deviations of observed phase
over various time scales, and (2) phase structure functions.  These
real-time statistical data measured by the API in the direction of the
satellite provide an estimate of the phase front distortion
experienced by the concurrent SMA astronomical observations.  The API
data also play an important role, along with the local opacity
measurements and weather predictions, in helping to plan the scheduling
of science observations on the telescope.

\end{abstract}

\maketitle

\section{Introduction}

Astronomical radio and millimeter interferometers combine signals from
multiple antennas in order to produce images of the sky.  The signals
must be combined with a predictable phase relationship based on the
array geometry and the direction of the target field
\citep{Ryle60}. In the vacuum of space, the signal propagates from an
astronomical source as a plane wave front. However, the atmosphere of
the Earth is not a homogeneous transmission medium, and spatial and
temporal variations in its constituents and properties along the line
of sight distort the original plane wave front \citep{Baars67}. The
resulting wave front aberration reduces the observed source visibility
as compared to the true source visibility, a phenomenon often termed
``radio seeing'' \citep{Baldwin90,Hargrave78}.  If the variations
become too large or too rapid, then they cannot be calibrated with the
standard technique of phase referencing in which astronomical point
sources are periodically observed to order to solve for the drift of
each antenna's complex gain vs. time
\citep[e.g.][]{Carilli99,Wright96,Shapiro79}.  In this situation, the
data will often become unusable, especially on the longer baselines
and at higher frequencies.  Therefore, in order to survey potential
sites for future interferometers and to promote optimal scheduling of
observations in different frequency bands on an existing
interferometer, it is desirable to construct an ancillary device to
measure atmospheric phase variations on a continuous basis.
Furthermore, these measurements should be performed on baseline
lengths comparable to those of the existing (or planned) array
configuration.


In this paper, we describe the design and operation of a novel,
broadband, multi-baseline atmospheric phase interferometer (API)
operating at the Submillimeter Array (SMA)\footnote{The Submillimeter
Array (SMA) is a collaborative project between the Smithsonian
Astrophysical Observatory and the Academia Sinica Institute of
Astronomy \& Astrophysics of Taiwan.} telescope site located near the
summit of Mauna Kea, Hawaii.  There are many examples of previous and
currently operating instruments that correlate geostationary satellite
beacon signals at astronomical observatory sites
\citep{Middelberg06,Hiriart02,Radford96,Ishiguro90,Masson90} and
elsewhere \citep{Kirkland00,Shao99}.  In contrast, rather than using a
beacon tone, our instrument correlates broadband digital television
emission from satellites which contains far more power than the beacon
tones, thereby enabling a high signal to noise ratio while avoiding
the many drawbacks of a very narrow bandpass filter .  The use of the
transponder signal rather than the satellite beacon to measure the
atmospheric phase is derived from the Berkeley Illinois Maryland Array
(BIMA) phase monitor design \citep{Lay98b}.  In addition to higher
signal to noise, the broadband system eliminates the possibility of
signal loss due to frequency drifts of the beacon tone that can exceed
the bandwidth of the narrow band filters required in beacon detection
systems.  The API acronym which we adopt in this paper for our system
is synonymous with the term Radio Seeing Monitor adopted by
\citet{Ishiguro94}.

\section{Instrument}

\subsection{Theory of operation} 

In the presence of a Gaussian distribution of atmospheric phase
distortions with standard deviation $\sigma$ (in radians at the
observing frequency), the observed visibility function of a celestial
source is reduced by an exponential factor: $V_{\rm obs} = V_{\rm
src}\exp{(-\sigma^{2}/2)}$ \citep{Thompson01}.  At submillimeter
wavelengths, the dominant atmospheric effect on the incoming wave
front is due to variations in tropospheric water vapor. The total
precipitable water vapor (PWV) toward the zenith above good sites for
(sub)millimeter astronomy typically ranges from $<1$ to several
millimeters. For reference, a variation of 0.1~mm PWV produces a 0.645
mm path length difference \citep{Shao99,Lay98a}, resulting in a phase
change of 4.0 radians at 300 GHz.  Fortunately, at microwave and
submillimeter wavelengths, the index of refraction of water vapor is
independent of frequency, except near strong absorption lines
\citep{Thompson01}. As a consequence, the atmosphere is mostly
non-dispersive, allowing one to extrapolate the phase front deviations
($\sigma_{\rm cm}$) measured at a low (centimeter wavelength)
frequency ($\nu_{\rm cm}$) into predicted phase deviations
($\sigma_{\rm submm}$) at submillimeter observing frequencies
($\nu_{\rm submm}$) by a simple ratio: $\sigma_{\rm submm} =
\sigma_{\rm cm} (\nu_{\rm submm}/\nu_{\rm cm})$.  With a typical
down-link frequency range of 11.7--12.7 GHz, K$_{\rm u}$ band
satellite digital television transmissions provide a convenient strong
signal with which to probe the tropospheric path variations.  We wish
to predict the phase deviations to an accuracy of 1\arcdeg\ at 700 GHz,
the highest frequency of operation of the SMA \citep{Hunter2005}.  In
order to achieve this performance using a 12~GHz API requires an rms
noise level of $\sim 3.5\times 10^{-4}$ radians.

\subsection{Description} 

\begin{figure}
\epsscale{1}
\plotone{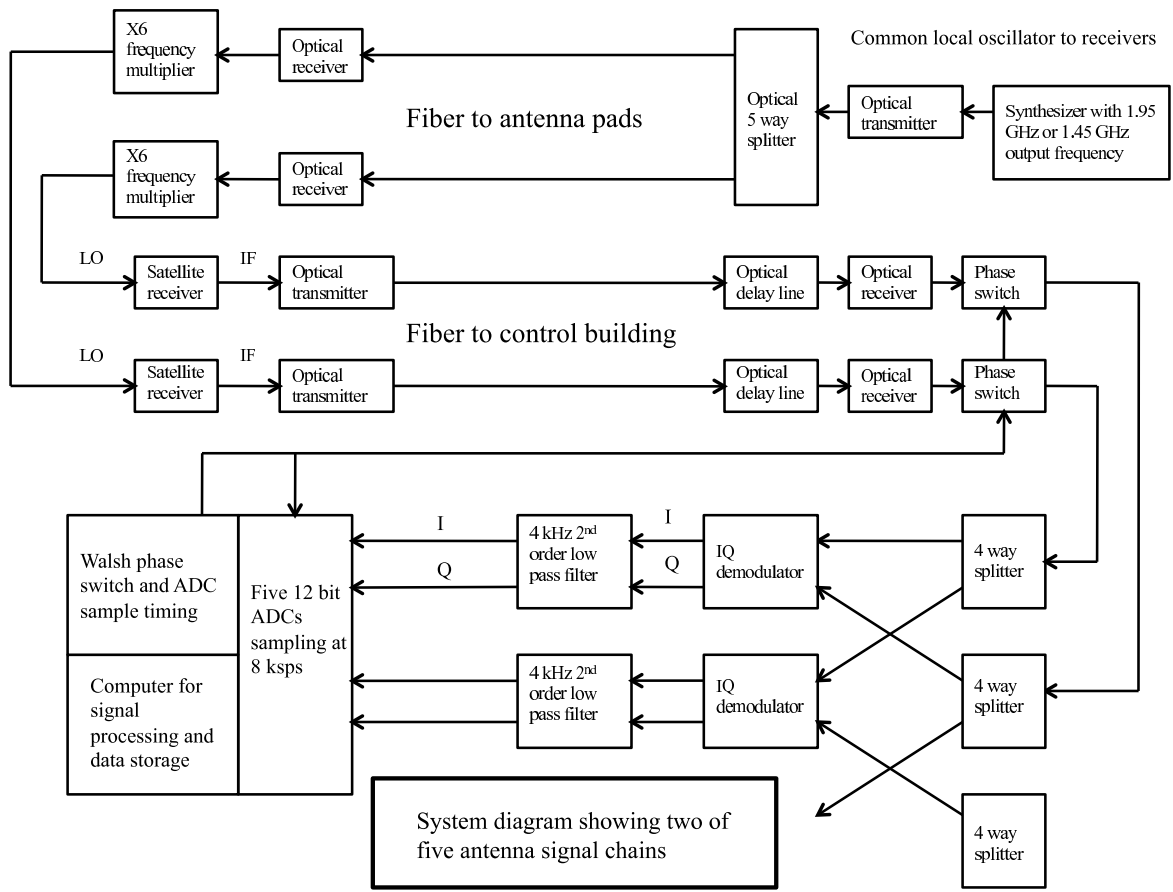}
\caption{Block diagram for two stations of the API and the associated IF electronics. \label{fig1}}
\end{figure}

\begin{figure}
\epsscale{1}
\plotone{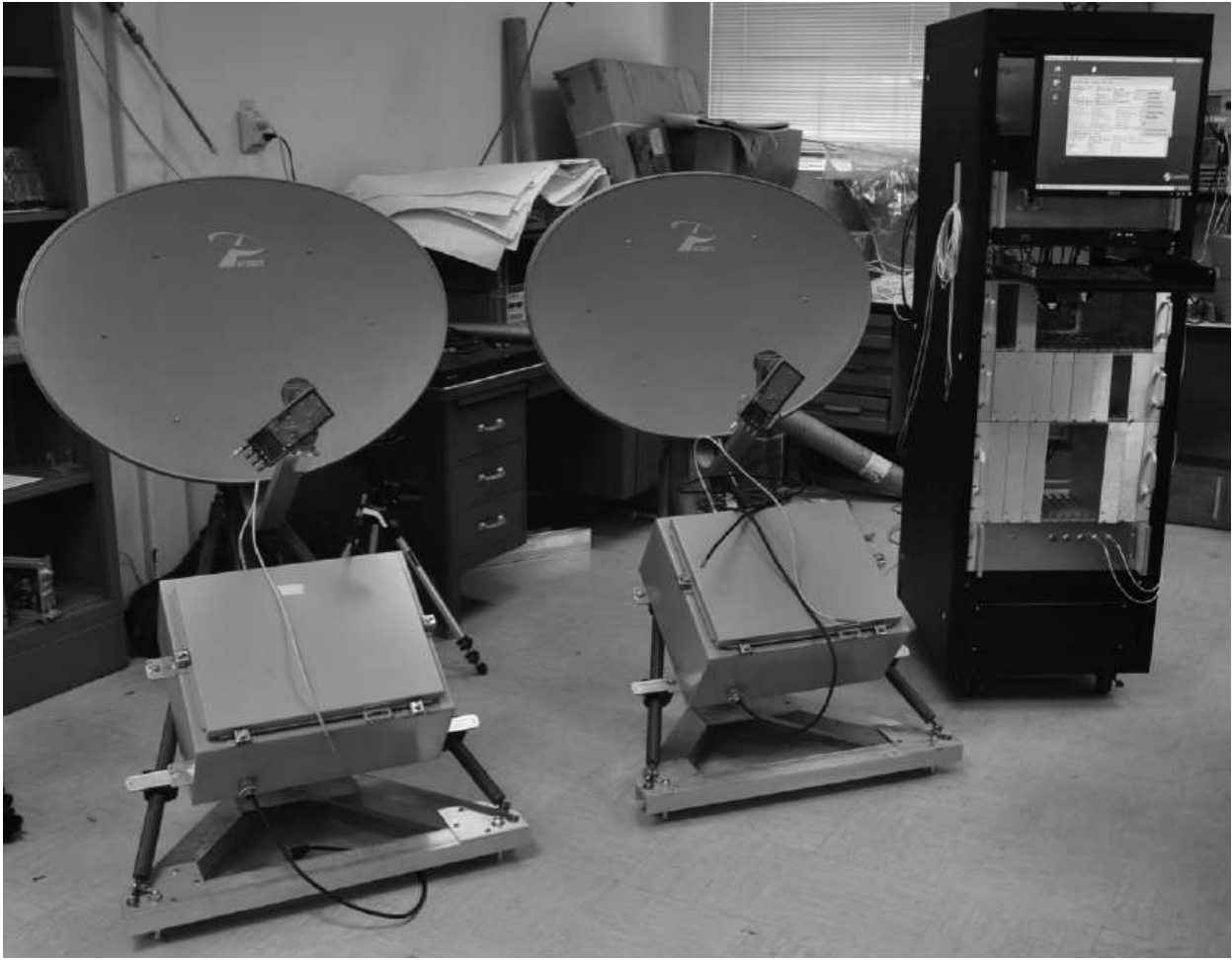}
\caption{Photograph of two stations of the SMA API and the central electronics rack
during laboratory testing.  The triangular base frame is bolted to the concrete
floor of the SMA antenna pad.
 \label{fig2}}
\end{figure}

The new API for the SMA is a multi-element 12 GHz interferometer that
observes a geostationary satellite.  A block diagram of the system is
shown in Figure~\ref{fig1}.  Each station consists of an inexpensive
(\$65) low noise mixing block (LNB) designed for satellite TV
reception coupled to a 0.7m diameter off-axis parabolic dish antenna
via a scalar feedhorn.  The LNB contains a low noise amplifier with a
rated noise temperature of 30~K (Invacom QPH-081), and we measured the
total receiver temperature to be 70~K.  We modified each LNB to accept
an external local oscillator (LO) signal coupled into a toroidal
dielectric resonator on the circuit board.\footnote{Larry D'Addario of
JPL developed a lower loss technique for injecting the LO into the LNB
while adapting the SMA API design for use by NASA's Deep Space Network
and as a site testing instrument for the Square Kilometer Array
(SKA).}  The LNB can simultaneously receive two linearly polarized
transponder bands and two circularly polarized transponder
bands. Separation is achieved by an orthomode transducer, and we use
the intermediate frequency (IF) signal from one of the circular
polarizations.  The digital quadrature phase modulated signal
transmitted by the satellite is spectrally white. The API's
correlation of a white noise signal has the benefit of eliminating any
spurious multipath effects \citep{Rogers93}. A white noise source of
60 MHz bandwidth will decorrelate in $\sim 5$~m of path difference so
only objects closer than that distance can potentially reflect an
interfering signal. However, the white noise correlation does impose
the need to equalize the delays from each antenna to the correlator
with spools of optical fiber.  A pair of assembled stations with their
NEMA enclosures is shown in Figure~\ref{fig2} along with the central
electronics rack.  Each station requires only 50 W of A/C power.

\begin{figure}
\epsscale{0.55}
\plotone{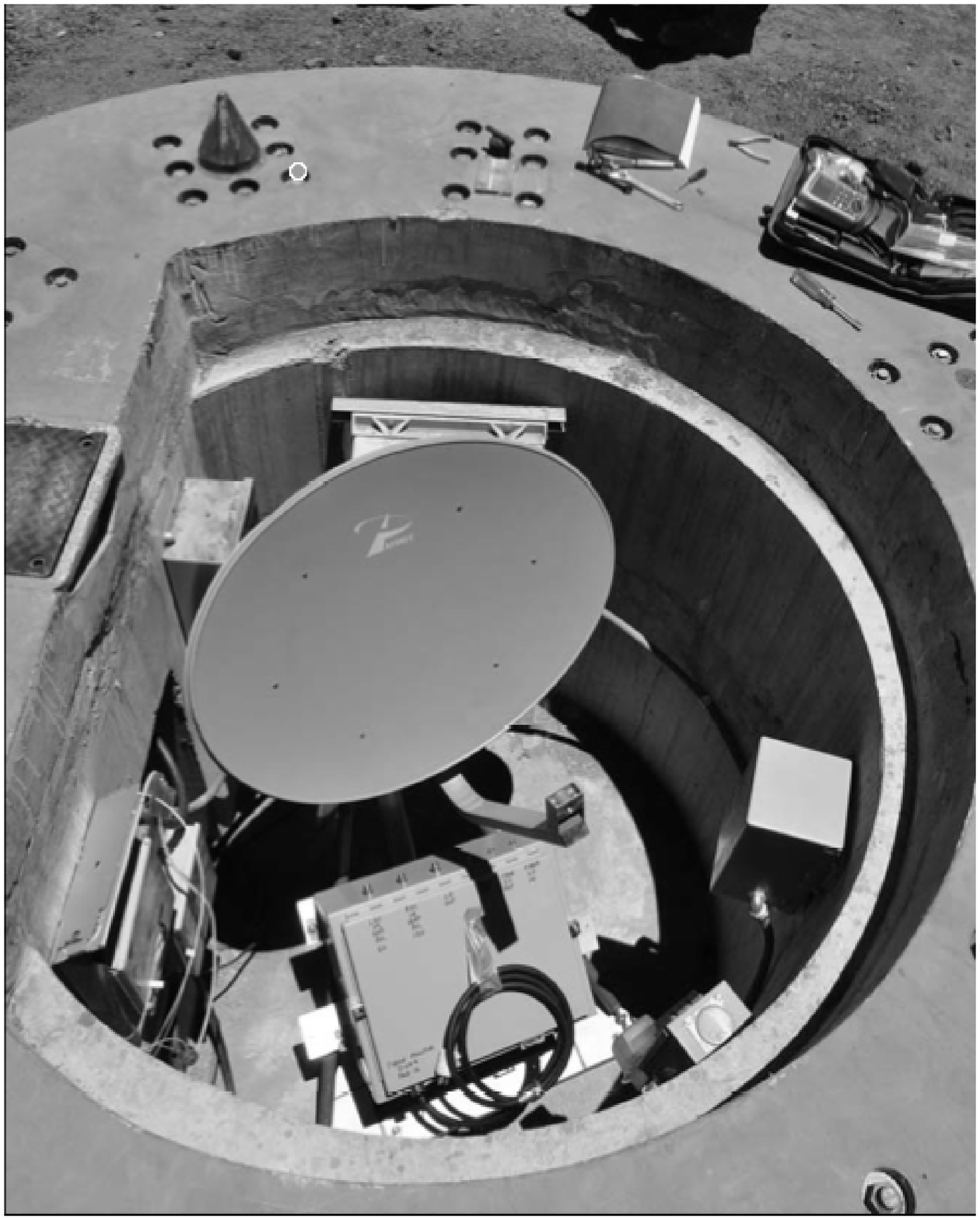}
\bigskip
\plotone{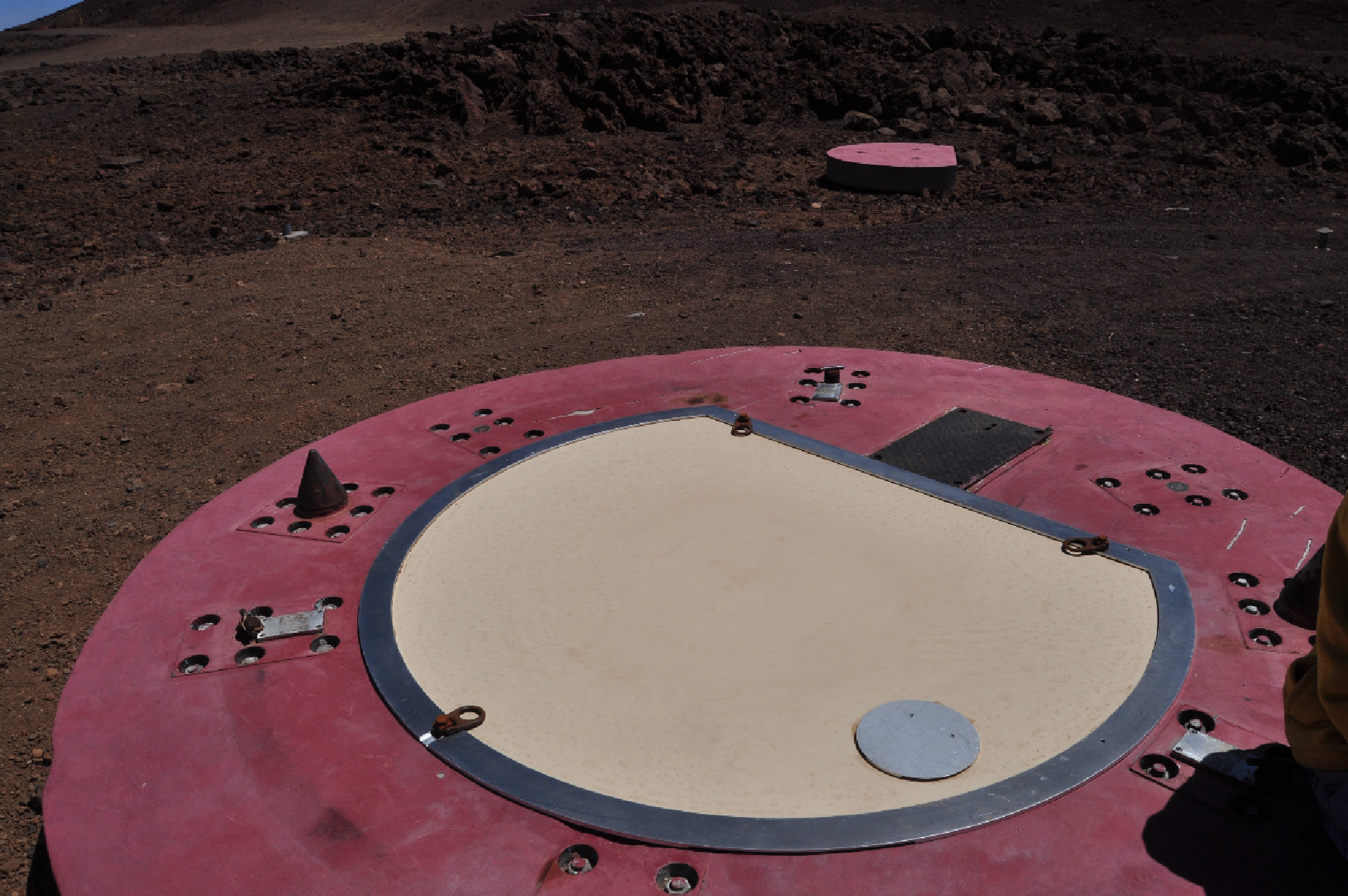}
\caption{Photographs of one station of the SMA API installed into a pad: (upper panel) with 
cover removed; (lower panel) with cover in place.  The short chimney provides 
ventilation by a small fan.
 \label{fig3}}
\end{figure}

The SMA site occasionally experiences significant periods of snow, ice
and high winds, which present a challenge for deploying sensitive
equipment.  However, because the SMA consists of 24 pads and 8
antennas, two thirds of the pads are unoccupied at any given time.
The idea of installing the API equipment in the pits of the unused
pads was conceived by former site director Antony Schinckel.  The
underground space containing the API antenna is hidden from view by an
environmental cover made from 19 mm thick marine grade high density
polyethylene. The cover prevents wind buffeting of the antennas, a
potential problem as the system is sensitive to micrometer scale
physical displacement of the antennas.  A photo of the dish antenna
and associated electronics box installed into an SMA pad is shown in
Figure~\ref{fig3}.  The station geometry of the API is listed in
Table~\ref{baselines}, including the pad numbers, the baseline lengths
(\bunprojected) and the projected lengths of the baselines
(\bprojected) as viewed from the target satellite.  A topographical
map of the SMA site with the pad locations indicated can be found in
\citet{Ho2004}, while further details of the SMA technical design,
digital phase-locked loops, and overall LO phase stability can be found in
\citet{Blundell2007}, \citet{Hunter2011}, and \citet{Kubo2006},
respectively.

Each of the antenna's electronics is connected via two 1330 nm
single-mode fiber optic cables to the equipment rack in the SMA fiber
patch panel vault located underground adjacent to the SMA control
building. The rack distributes the common LO signal on one fiber and
collects the intermediate frequency (IF) signals from the other fiber.
These fibers are two of the three available fibers that can be used to
carry the submillimeter receiver LO and IF signals when an SMA antenna
is present on the pad.
The use of these buried fiber cables (which were custom designed to
have a low phase vs. temperature coefficient) combined with the
placement of the central electronics in an underground room promotes a
phase-stable system by effectively eliminating phase drifts due to
diurnal temperature swings.  The IF signals from each station are
processed as pairs on ten ``baseline boards''.  Each board contains a
CMOS analog switch driving the IF port of a mixer which applies an
alternating 180\arcdeg\ phase shift to one of the two station signals.
The signal enters the RF port of the mixer and emerges from the LO
port before proceeding to the analog correlator.  This Walsh pattern
modulation \citep{Granlund78} removes the 1/f noise (drift) of the
correlator's Gilbert cell multipliers, following the example of the
BIMA phase monitor \citep{Lay98b}.  The IF signal pairs are processed
into in-phase ($I$) and quadrature ($Q$) outputs by the correlator
which is a broadband direct conversion quadrature ($I/Q$) demodulator
(Analog Devices AD8347) commonly used in digital satellite television
receivers.  These baseline pair outputs are digitized at a rate of 8
kHz by an analog to digital converter card on the PCI bus of a
dedicated Linux computer.  The baseline phases and amplitudes are
calculated from $\arctan{(Q/I)}$ and $\sqrt{Q^2+I^2}$, respectively.
The Walsh demodulation is performed in software by the API computer
control program, which subsequently averages the data to 10 Hz.

\subsection{Satellite target and sensitivity}

\begin{figure}
\epsscale{1}
\plotone{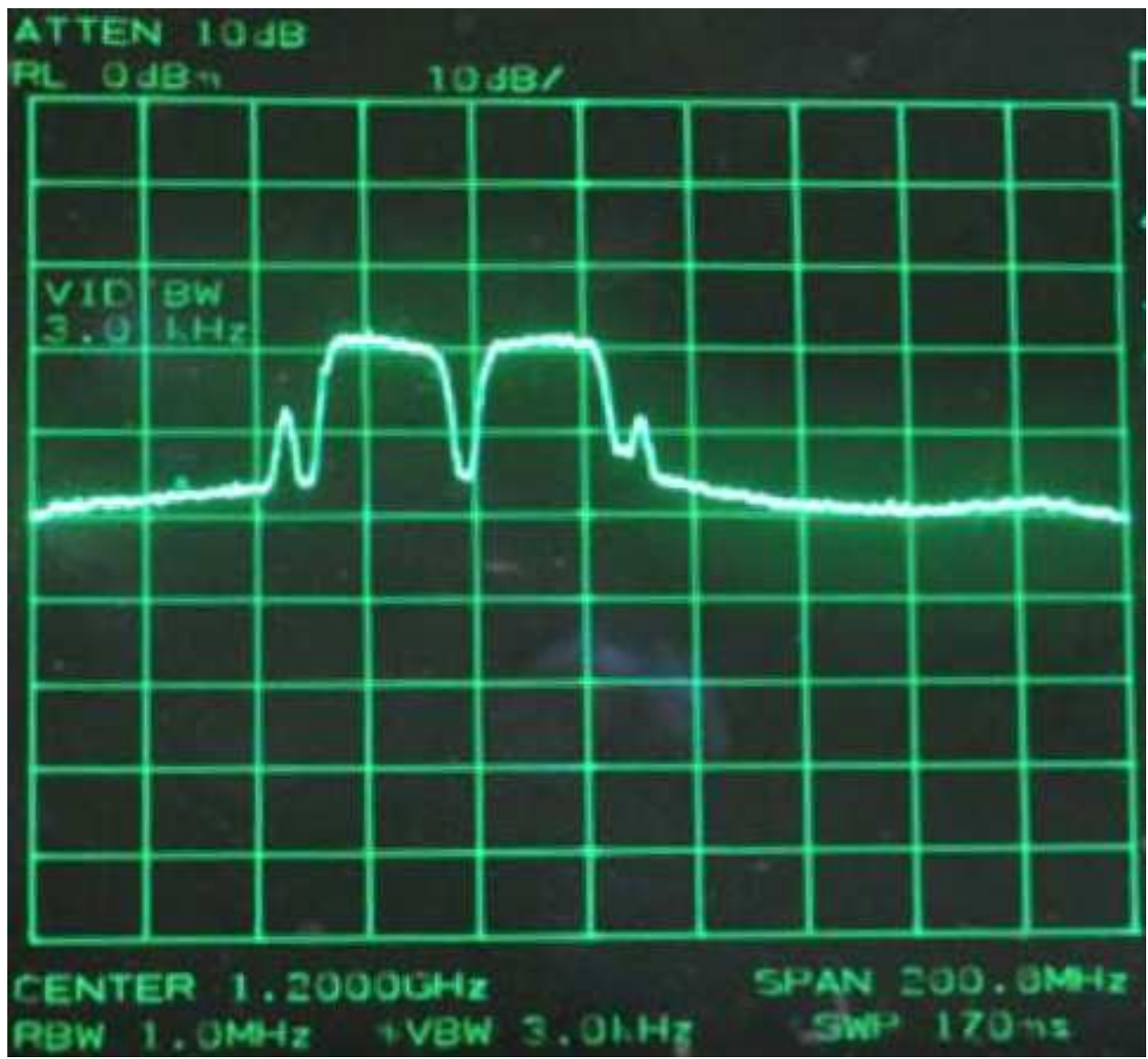}
\caption{Spectrum analyzer trace of the IF output of one station of the API with
the dish antenna pointed at the satellite Ciel-2.  In this case, the signal power
to noise power ratio exceeds 10 dB, but the typical value is $\geq 8$ dB.
 \label{fig4}}
\end{figure}

The satellite we have chosen to observe is Ciel 2 (NORAD 33453). This
Canadian satellite is a Spacebus-4000C4 model launched in 2008 and
currently operating at 129\arcdeg\ West geostationary orbit position.
From Mauna Kea, it appears at an azimuth of 124\arcdeg\ and an
elevation of 52\arcdeg\ which is sufficiently high to avoid vignetting
of the beam by the cylindrical wall of the concrete pad.  The
equivalent isotropic radiated power (EIRP) is 52~dBW in each of its
30~MHz transponder bands.  A spectrum analyzer trace showing the
satellite signal bands at 12.4~GHz used by the API correlator is shown
in Figure~\ref{fig4}.  The signal to noise ratio ($SNR$) of the received
satellite signal power is about 8~dB, which places the API system in the 
``strong source'' case when considering the sensitivity.
%
For our purposes, we are interested in the theoretical 
noise floor of the API with respect to the measured fringe phase angle.
We consider the real signal and noise voltages from two antennas to be
$S_1(t)$, $S_2(t)$, $N_1(t)$, and $N_2(t)$.  With the delay set such
that the satellite is at the center of the white light fringe, $S_1(t)
= S_2(t) \equiv S(t)$.  Consequently, the $I$ output from the $I/Q$
demodulator will be proportional to $(\langle S(t)\rangle_{\rm rms})^2
\equiv {\srms}$ because only this term integrates coherently and the
other terms will be small.  In the $Q$ output, the $\srms$ term will
integrate to zero due to the 90\arcdeg\ phase shift, thus we are left
with the sum of the cross terms and the noise product term.  Since
$(\langle N_1\rangle_{\rm rms})^2 \approx (\langle N_2\rangle_{\rm
rms})^2 \equiv N_{\rm rms}^2$, the rms deviation of $Q$ from zero 
will follow the radiometer equation for the case of
identical receivers and antennas (see Eq. 7-32 of \citet{Crane1989}).
Therefore, we can relate the theoretical rms noise
floor of the API in terms of the rms fringe phase angle, $\sigma_{\rm rms}$ 
(in radians), to the $SNR$, the received bandwidth ($B$), and the 
integration time ($\tau$) by starting with the small angle approximation:
\begin{equation}
\sigma_{\rm rms} = \langle\arctan{(Q/I)}\rangle_{\rm rms} \approx Q_{\rm rms}/I = \frac{\sqrt{(2\srms\nrms+\nrmsf)/(B\tau)}}{\srms}  = \sqrt{\frac{2/SNR + 1/SNR^2}{B\tau}}. 
\end{equation}
In cases such as ours where the $SNR$ is large, a further approximation can be made:
\begin{equation}
   \sigma_{\rm rms} \approx \sqrt{\frac{2/SNR}{B\tau}}, 
\end{equation}
which is different from the approximation applied in the more typical, ``weak 
source'' radio astronomy case.
%
%
With the current operating parameters of $B = 60 \times 10^6$ Hz, and
$t = 0.1$~s, the theoretical noise floor is $1.3 \times 10^{-4}$
radians or 0.013\arcdeg\ at 12.4 GHz. The practical noise level
achieved in the laboratory, 0.02\arcdeg\ rms, is only about 50\% worse
than this prediction.  At 700 GHz, the measured noise corresponds to
an rms noise floor of 0.73\arcdeg\ phase, which meets the requirement.

\subsection{Data products}

Because geostationary satellites cannot maintain a perfectly circular
orbit exactly in the equatorial plane, they typically exhibit a small
apparent diurnal motion when viewed from the Earth.  This motion
amounts to several turns of phase on our baselines.  We remove this
effect in real time by automatically computing a running sinusoidal
least-squares fit to the phase data.  The standard deviation of the
phase on each baseline is then computed for a series of time scales
ranging from 1 to 2048 seconds in powers of two.  The square root of
the phase structure function \citep{Tatarski61} is also computed using
these same time scales plus the additional lags of 768, 1536, and 2816
seconds. The processed data are then written to the SMA distributed
shared memory (DSM) system, which allows them to be displayed on the
operator's monitor screens and automatically written to the central
engineering data archive.  The cross-correlation amplitudes from each
baseline and the auto-correlation amplitudes from each station are
also produced by the demodulators as voltage outputs which are
sampled, copied to DSM, and stored to the engineering data archive.
These amplitude values can alert the operators to the presence of
water or snow on the environmental shield as it can substantially (or
completely) attenuate the satellite signal.  When necessary,
functionality can then be restored by sweeping the shields clear.

\subsection{Calibration of the I/Q demodulator}

\begin{figure}
\epsscale{1}
\plotone{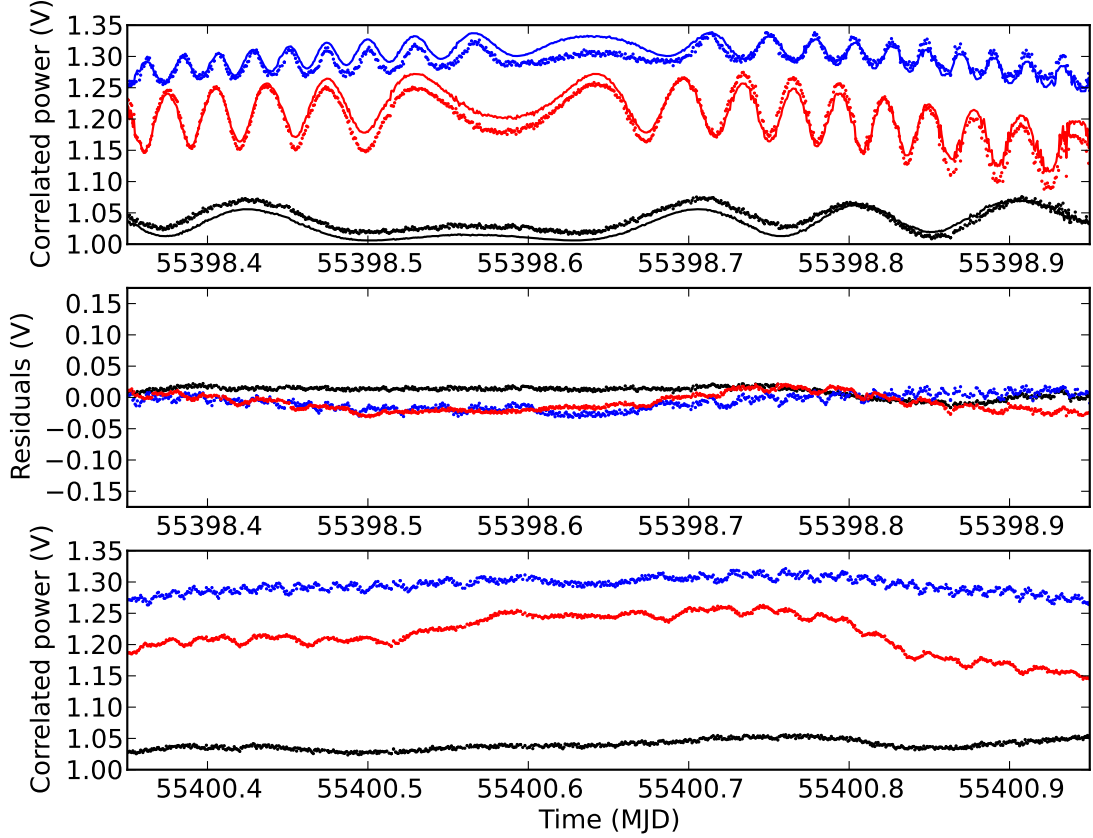} 
\caption{The baseline correlated power vs. time for three API baselines. 
Dots are the measurements and lines are the model fits for the relative
gains of the $I/Q$ demodulator outputs.  Panel (a) shows the raw data 
with the model fit overlaid, from top to bottom the data are from 
baselines: 2--4 (\bprojected\ = 86m), 3--4 (122m), and 1--2 (33m).  
Panel (b) shows the residuals after model subtraction. Panel (c) shows 
the raw data from two days later, after the model 
coefficients were installed into the software.  The high frequency 
variation is mostly eliminated.  
 \label{fig5}}
\end{figure}

After the initial mechanical attachment and manual alignment of the
dish antennas in each pad, the instrument requires very little in
terms of calibration.  We did find that the $I/Q$ demodulators exhibit
a small difference in the complex gain between their two output
channels.  Because of the satellite's diurnal motion, this imbalance
leads to a predictable pattern in the cross-correlation amplitude
vs. time, typically about 5\% in terms of the peak to peak variation
with $\approx$ 0.5--3 dozen periods per day, depending on the baseline
length. We performed a least-squares fit to this pattern over several
days of data from each baseline and derived the appropriate software
gain factors for each demodulator in order to bring the outputs into
agreement.  After applying these gain factors to the digitized
voltages in software, a large fraction of the amplitude pattern is
eliminated (Figure~\ref{fig5}).

\section{Results}

The first three API stations were installed in late 2009, the fourth
in July 2010, and the fifth and final station in July 2011.  In this
section, we present preliminary results from a two month period from
November 1 through December 31, 2011 during which the SMA was in
compact configuration, enabling all stations to see the sky
simultaneously.  Nine of the ten baseline boards were functioning
properly throughout this period. Baseline 4-5 had consistently low
correlated amplitude due to a fiber delay error, so those data are
ignored in the following analysis. The delay error was corrected on
January 11, 2012.

\subsection{Examples of variable atmospheric conditions in the API data}

\begin{figure} 
\epsscale{1}
\plotone{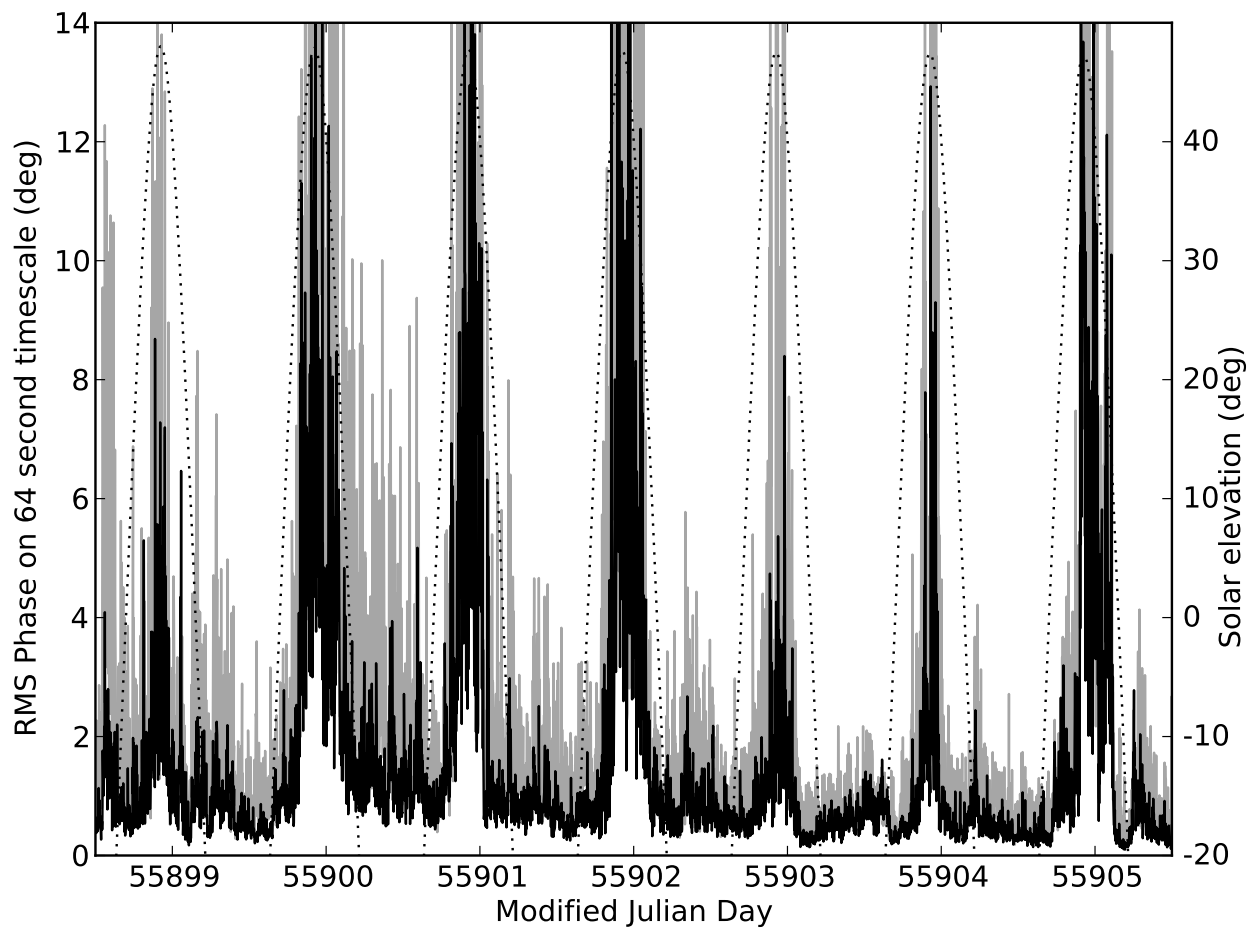}
\caption{The rms phase at 12.4~GHz on 64-second intervals measured on two different
baselines (black line: shortest baseline (\bprojected\ = 33~m), gray 
line: longest baseline (\bprojected\ = 261~m) over a one 
week period in early December 2011.  The dotted line is the 
elevation of the sun, as specified on the right side y-axis.  The diurnal 
cycle of lower and more stable rms phase at nighttime is clearly evident.
 \label{fig6}}
\end{figure}

\begin{figure}
\epsscale{1} 
\plotone{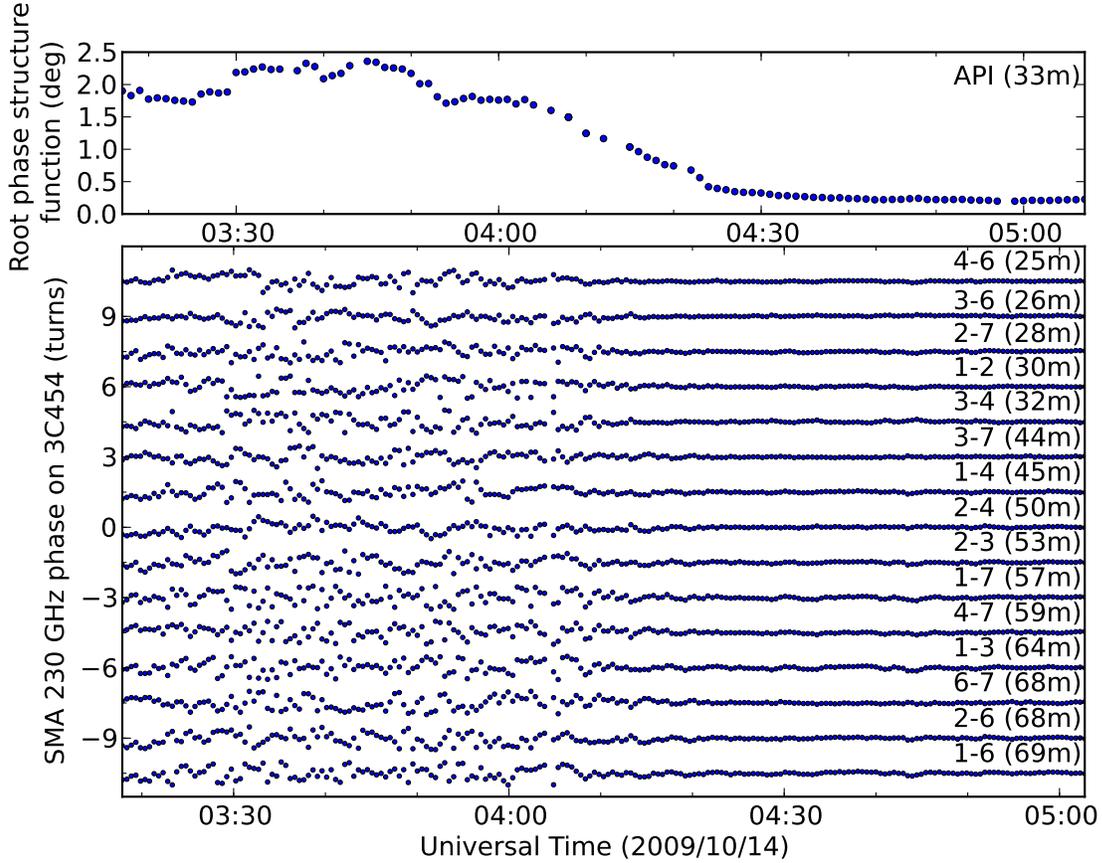}
\caption{Top panel: The 32 second root phase structure function from
the shortest API baseline.  Bottom panel: The calibrated SMA interferometer 
phase recorded on 15 baselines while observing the bright quasar 3C454 with
a 2 GHz bandwidth centered at 230.25 GHz with 30 second integrations.
The SMA antenna pairs and the corresponding unprojected baseline lengths 
are listed above each trace, arranged in order of length.  The traces 
are successively displaced by 1.5 turns in order to make it easier to 
distinguish between the data points.
 \label{apisma}}
\end{figure}

The rms phase from two selected baselines of the API (the shortest and
longest) over a one week period is shown in Figure~\ref{fig6}.  The
well-known diurnal cycle of good phase stability above Mauna Kea at
night and poor stability during the day \citep{Masson94a} is clearly
evident in the data.  One can also see that on some nights the
atmosphere is significantly more stable than on others.  The data from
the different baselines behave similarly but not identically, which is
expected given the different baseline lengths and orientations and the
nature of water vapor fluctuations.  The rms is generally worse on
longer baselines but by varying amounts.  Detailed comparisons of the
variation in the rms phase measured by the API with the SMA
interferometer phase stability measured on calibrators typically show
excellent agreement.  To illustrate this agreement quantitatively, we
continuously observed a strong quasar with the SMA in compact
configuration over a two hour interval during which the observing
conditions improved drastically.  Figure~\ref{apisma} compares the
32-second time scale root phase structure function from the shortest
API baseline with the 230 GHz phase recorded at 30 second integrations
from each SMA baseline.  The median of the per-baseline standard
deviations of the 230 GHz phases dropped by a factor of 9.5, from
103\arcdeg\ during the 03:30-04:00 period to 10.6\arcdeg\ during the 
04:30-05:00 period.  Over these same periods, the standard deviation
of the API root phase structure function dropped by a factor of 9.7,
from 2.18\arcdeg\ to 0.23\arcdeg.  This agreement demonstrates that
the API data provide a full-time, direct assessment of the
submillimeter observing conditions on Mauna Kea for comparable
interferometric baselines.

\subsection{Comparison of rms phase among the API baselines}

\begin{figure}
\epsscale{1} 
\plotone{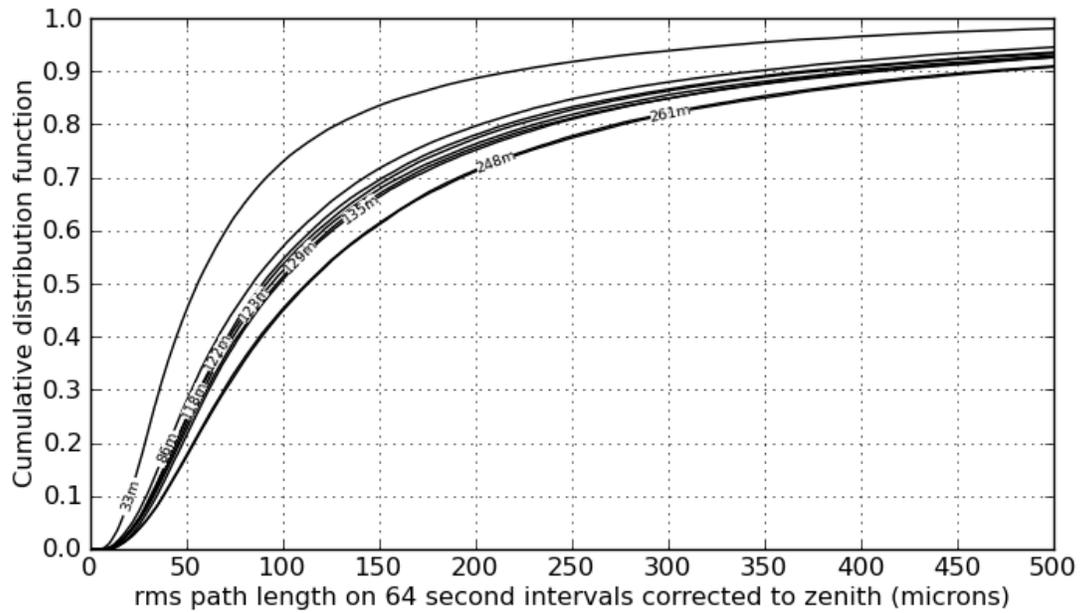}
\caption{The 24-hour cumulative distribution functions of the rms
path length (corrected to zenith) on 64 second intervals 
on nine of the API baselines, labeled by their length (\bprojected).
 \label{rmscdf}}
\end{figure}

The cumulative distribution function of the rms path length on a 64
second timescale is shown in Figure~\ref{rmscdf} for all nine API
baselines corrected to the zenith by dividing by $\sqrt{\rm
sec(zenith~angle)}$ \citep{Coulman85}.  There is an extended tail in
the distribution toward higher values at all timescales, reflecting
the worst observing conditions on Mauna Kea.  Therefore, rather than
using the mean and standard deviation, we compute the median and
median absolute deviation (MAD) for the zenith-corrected rms path
length in order to better represent the variation in the majority of
the data.  Listed in Table~\ref{percentiles} are the results for the
512 second timescale which provides an estimate of the median rms
phase variation between typical visits to a phase calibrator during a
science observation.  Considering that the medians are derived from
only two months of data, the nighttime values agree well with the
overall nighttime median value of $101 \mu$m for Mauna Kea measured
over 1406 days from 1990-1998 with the original SAO 100~m baseline
phase monitor \citep{Moran98}.  For our calculations, ``nighttime'' is
an approximate term that refers to the period of time from sunset plus
two hours to sunrise plus two hours, and is quite similar to the
definition used by \citet{Moran98} of 20 hours to 8 hours local time
for this tropical location.  Finally, the smaller values of both the
median and the MAD found during nighttime are further indication of
the diurnal effect.

\subsection{Root phase structure function slope and corner time}

\begin{figure}  
\epsscale{1}
\plotone{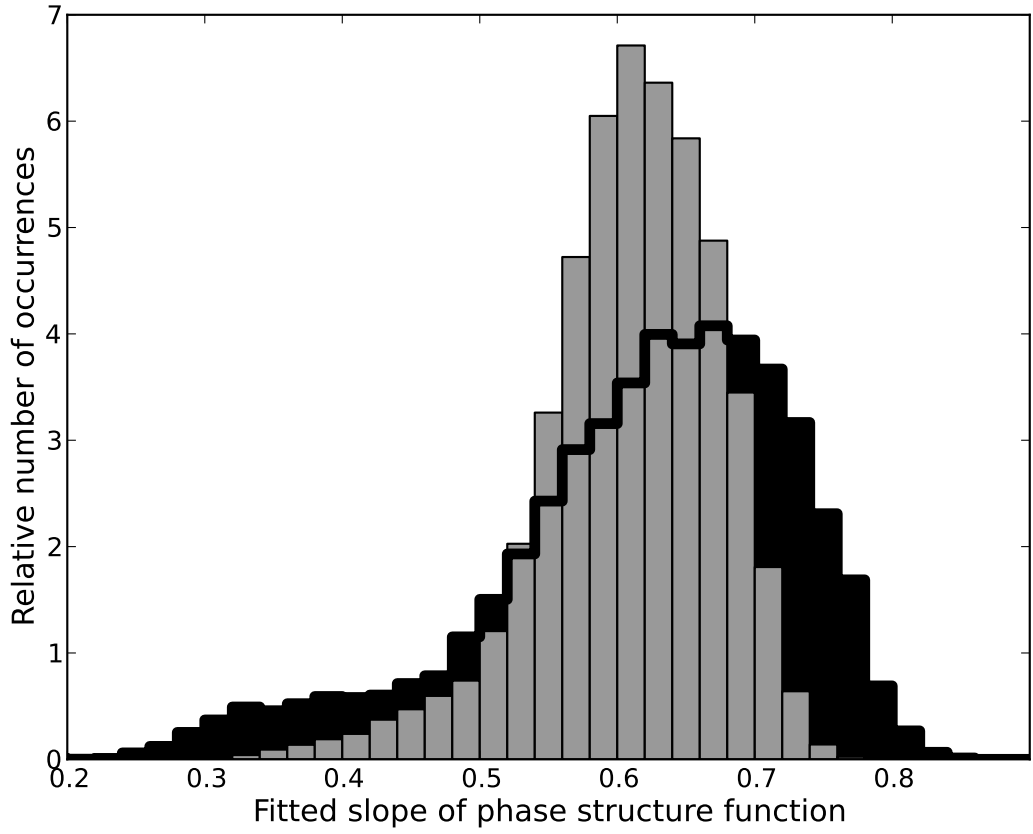} 
\caption{Histograms of the fitted slope of the phase structure
function ($\alpha$) for the shortest baseline (\bprojected\ = 33m) in black,
and the longest baseline (\bprojected\ = 261m) in gray. 
 \label{alphahistograms}} 
\end{figure}

\begin{figure}  
\epsscale{1}
\plotone{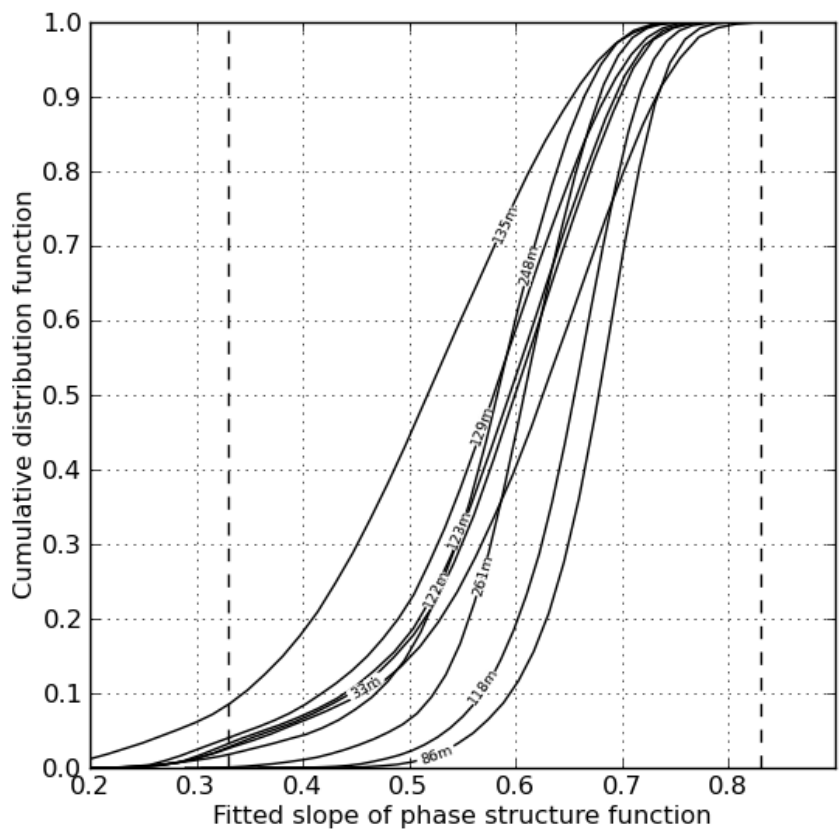}
\caption{Cumulative 
distribution functions of the fitted slope of the phase structure
function $\alpha$ for nine baselines labeled by \bprojected. 
The dashed vertical lines mark the expected Kolmogorov values for 
thin atmospheres (0.33) and thick (0.83) atmospheres. \label{alphacdfs}} 
\end{figure}

Each baseline of the API provides an independent measurement of the
root phase structure function.  For each baseline, we have fit the
slope ($\alpha$) and corner time (\tcorner) of the root phase
structure function for each time interval following the iterative
procedure of \citet{Holdaway95}.  To find the initial slope, we
perform a linear least squares fit to the first six data points (lags
of 1-32 seconds).  We derive the initial corner time by computing
median value of the structure function over the longer lags (beginning
at 64 seconds) and find the time where the linear fit crosses this
value.  This process is repeated until convergence, which requires at
most a few iterations.  The median and MAD values of $\alpha$ and
\tcorner\ per baseline over the two month period are listed in
Table~\ref{percentiles}.  The histogram of values of $\alpha$ for the
shortest and longest baselines and a cumulative distribution function
(CDF) for all nine baselines are shown in
Figures~\ref{alphahistograms} and \ref{alphacdfs}.  In this table and
figures, we have excluded data from the time intervals when any of the
baselines reported abnormally low normal amplitudes, that is, when one
or more of the pad covers were blanketed with snow.  As seen with the
previous SAO instrument on Mauna Kea \citep{Masson94a}, $\alpha$
varies over a range of values (0.2-0.8).  The median value of $\sim
0.65$ is intermediate between the expected Kolmogorov exponent for a
thick atmosphere (0.83) and a thin atmosphere (0.33)
\citep{Thompson01}.  The relatively small values of the MAD
demonstrate that this intermediate value of $\alpha$ is persistent,
and does not represent a time average between the two extreme cases.
In fact, intermediate values for $\alpha$ of $\sim 0.7$ are seen at
many other astronomical telescope sites \citep{Masson94b}.  The corner
times of $\sim 200$ sec correspond to a flattening of the frequency
spectrum of phase variations below $\sim 0.005$ Hz, but with a fairly
large variation.  Again, this result is consistent with the
conclusions from the previous SAO phase monitor on Mauna Kea.
Finally, we note that the shapes of the histograms and CDFs are very
similar to those derived for the ALMA site from the NRAO site test
interferometer \citep{Holdaway95,ALMA}.

\subsection{Practical usage of the API}

During each six-month semester, the SMA normally cycles through four
configurations ranging from sub-compact to very-extended
\citep{Petitpas2010}.  The range of baselines is given in
Table~\ref{configs}.  In the extended configurations, some of the API
antennas are covered by SMA antennas; however, in all cases at least
three API baselines are operational with projected lengths ranging
from 33~m to at least 129~m.  The API data have become a useful tool
for the observatory's science operations team.  The long term use of
the API continues to emphasize the fact that the phase stability is
generally worse during the day and better at night.  For this reason,
the SMA typically executes science projects from very late afternoon
until some number of hours after sunrise \citep{Christensen2008}.  The
superior nighttime phase stability generally persists for some period
after sunrise (see Fig~\ref{fig6}), but the length of this period is
quite variable.  However, the deterioration in phase stability is
easily seen in the API data, and by watching it closely, the current
science observation can be curtailed when necessary by acquiring the
remaining calibration observations before the conditions become
unusable.  Unfortunately, the degree to which the day vs. night phase
stability dichotomy holds true does not seem to be correlated with any
other better studied atmospheric variables such as PWV (see also
\citet{Masson94a}), relative humidity, etc. Therefore, the use of the
API as a predictive tool for scheduling upcoming science observations
is limited, and those decisions are still generally dominated by the
PWV forecast.

In addition to tracking the diurnal conditions, the API data helps to
alert the operations team to the passage of major weather patterns.
For example, after just a few months of use, it became clear that poor
phase stability was almost always assured when the forecast predicted
a transition from very dry to very wet conditions. It is also
occasionally true that transitions from very wet to very dry
conditions also result in poor phase, but not always (in contrast to
the case for the reverse trend). Therefore, if we are expecting half
of the night to be wet (or dry) changing to dry (or wet), it is almost
always prudent to schedule a lower frequency project, because even if
the opacity improves, one can expect the phase to remain unstable for
longer.  Finally, we do occasionally see long term trends of poor
phase stability at night, typically when there is a lot of moist air
moving across the Hawaiian Islands. During these periods, we tend to
preferentially schedule lower frequency projects since they are less
affected by poor phase stability than the higher frequency projects
and have a greater chance of success. We can then monitor the API data
during the day to watch for improvements in phase stability which
could suggest that it is worth trying higher frequency observations that
evening.

\section{Summary}

We have successfully developed and deployed a five-station 12~GHz API
for the SMA. Each of the ten baselines correlates a broadband digital
signal from a geostationary television broadcast satellite.  The
stations are installed in the covered pits of SMA antenna pads,
protecting them from environmental disturbances, and yielding a path
length sensitivity of $\approx 1$ micron.  The real-time phase
stability data correlates well with the observed phase stability of
the SMA interferometer data stream.  Preliminary statistics on the rms
path length and the behavior of the root phase structure function
(slope and corner time) are consistent with the previous single
baseline atmospheric phase monitor experiment on Mauna Kea.  The SMA
API is used regularly along with other measurements of weather
conditions and forecasts to schedule science observations in
appropriate conditions for the desired frequency band, and to help
ensure that these observations are completed successfully.  Future
studies of the phase stability statistics for Mauna Kea will become
possible as the archive of API data grows.

\acknowledgments

We thank John Kovac for assistance in the measurement of the receiver
temperature of the LNB, and Jim Moran for useful discussions.  The
National Radio Astronomy Observatory is a facility of the National
Science Foundation operated under cooperative agreement by Associated
Universities, Inc.  This research has made use of NASA's Astrophysics
Data System Bibliographic Services.  The authors extend special thanks
to those of Hawaiian ancestry on whose sacred mountain we are
privileged to be guests.

\begin{table}
\caption{Stations and baseline lengths in the SMA API.  
\label{baselines}}
\begin{tabular}{ccccccc} 
\hline
\multicolumn{2}{c}{Baseline} & \multicolumn{2}{c}{Baseline length (m)} & Pad height & Azimuth & Blocked in\\
Stations & pads & \bunprojected  & \bprojected & difference (m) & (\arcdeg) & configuration\tablenotemark{a}\\
\hline
1-2 & 22-10 &  46 & 33  &  6.7 & 143/323 & -- \\
2-3 & 10-14 & 134 & 123 &  4.8 &  72/252 & ext\\
3-1 & 14-22 & 156 & 129 & 11.4 &  88/268 & ext\\
4-1 & 15-22 & 158 & 118 & 11.8 & 134/314 & ext\\
3-4 & 14-15 & 124 & 122 &  0.3 &  20/200 & ext\\
2-4 & 10-15 & 114 & 86  &  5.1 & 131/311 & ext\\
\hline 
2-5 & 10-18 & 299 & 248 & 18.0 &  90/270 & vex\\
3-5 & 14-18 & 176 & 135 & 13.2 & 103/283 & vex\\
4-5 & 15-18 & 227 & 206 & 12.9 &  70/250 & vex\\
5-1 & 18-22 & 328 & 261 & 24.7 &  96/276 & vex\\
\hline
\end{tabular}
\tablenotetext{a}{In extended and very-extended configurations, some of the baselines are blocked 
by SMA antennas.  Baseline 1-2 is never blocked because pads 10 and 22 are never occupied by SMA antennas in the standard sequence of configurations.}
\end{table} 

\begin{table}
\caption{Baseline lengths of SMA and API during the various SMA configurations\label{configs}}
\tablenotetext{a}{Projected baseline lengths perpendicular to the target
satellite direction}
\begin{tabular}{cccc}
\hline
Configuration & Range of SMA   & Number of & Range of API \\
of SMA        & baselines (m)  & stations  & baselines (m)\tablenotemark{a}\\
\hline
sub-compact   &  9.5-69 & 5 & 33-261\\
compact       &  17-77  & 5 & 33-261\\
compact-north &  17-139 & 5 & 33-261\\
extended      &  44-227 & 3 & 33-261\\
very-extended &  68-508 & 4 & 33-129\\
\hline
\end{tabular}
\end{table}

\begin{table}
\caption{Median and median absolute deviation (MAD) of $\alpha$, \tcorner, and 
rms path length for nine API baselines\label{percentiles}} 
\begin{tabular}{cccccc}  
\hline
         &               &              &                 &  \multicolumn{2}{c}{rms path length (microns)\tablenotemark{a}} \\
Baseline & Length (m)    & $\alpha$     & \tcorner\ (sec) & 24-hour      & Nighttime\tablenotemark{b} \\
 (pads)  & \bprojected\  & Median (MAD) & Median (MAD)    & Median (MAD) & Median (MAD)\\
\hline
22-10 & 33   & 0.63 (0.07) &  92  (35) &  73  (38) &  60 (28) \\
10-14 & 86   & 0.69 (0.03) & 155  (68) & 127  (71) & 104 (51) \\ 
14-22 & 118  & 0.66 (0.04) & 193  (95) & 145  (82) & 119 (59) \\
15-22 & 122  & 0.61 (0.06) & 203 (101) & 135  (73) & 111 (52) \\
14-15 & 123  & 0.60 (0.06) & 221 (109) & 144  (81) & 118 (57) \\
10-15 & 129  & 0.59 (0.06) & 246 (123) & 150  (85) & 122 (60) \\
18-22 & 135  & 0.53 (0.08) & 299 (142) & 153  (87) & 126 (62) \\
14-18 & 248  & 0.59 (0.05) & 359 (175) & 193 (108) & 160 (78) \\ 
15-18 & 261  & 0.62 (0.04) & 347 (170) & 198 (110) & 163 (79) \\
\hline
\end{tabular}
\tablenotetext{a}{Calculated over 512 second intervals and corrected to the zenith.}
\tablenotetext{b}{Nighttime is defined as sunset plus 2 hours to sunrise plus 2 hours.}
\end{table}

\clearpage

\end{document}